\documentclass[twocolumn,english,superscriptaddress,citeautoscript,showpacs,preprintnumbers,amsmath,amssymb,prl,floatfix,footinbib]{revtex4}
\usepackage[scaled=2]{helvet}
\usepackage[T1]{fontenc}
\usepackage[latin9]{inputenc}
\usepackage{array}
\usepackage{multirow}
\usepackage{amsmath}
\usepackage{amssymb}
\usepackage{graphicx}

\makeatletter

\providecommand{\tabularnewline}{\\}

\@ifundefined{textcolor}{}
{%
 \definecolor{BLACK}{gray}{0}
 \definecolor{WHITE}{gray}{1}
 \definecolor{RED}{rgb}{1,0,0}
 \definecolor{GREEN}{rgb}{0,1,0}
 \definecolor{BLUE}{rgb}{0,0,1}
 \definecolor{CYAN}{cmyk}{1,0,0,0}
 \definecolor{MAGENTA}{cmyk}{0,1,0,0}
 \definecolor{YELLOW}{cmyk}{0,0,1,0}
}

\usepackage[scaled=2]{helvet}\usepackage{amstext}

\makeatletter
\@ifundefined{textcolor}{}{\definecolor{BLACK}{gray}{0}\definecolor{WHITE}{gray}{1}\definecolor{RED}{rgb}{1,0,0}\definecolor{GREEN}{rgb}{0,1,0}\definecolor{BLUE}{rgb}{0,0,1}\definecolor{CYAN}{cmyk}{1,0,0,0}\definecolor{MAGENTA}{cmyk}{0,1,0,0}\definecolor{YELLOW}{cmyk}{0,0,1,0}}

\usepackage{dcolumn}
\usepackage{bm}
\usepackage{times}
\usepackage{hyperref}
\hypersetup{colorlinks=true,linkcolor=red,citecolor=blue}
\makeatother

\usepackage{babel}

\makeatother

\usepackage{babel}
\begin{document}

\title{Magnetic structure of the Eu$^{2+}$ moments in superconducting EuFe$_{2}$(As$_{1-x}$P$_{x}$)$_{2}$
with \emph{x}\,=\,0.19}

\author{S. Nandi}

\email{s.nandi@fz-juelich.de}

\selectlanguage{english}%

\affiliation{Jülich Centre for Neutron Science JCNS and Peter Grünberg Institut
PGI, JARA-FIT, Forschungszentrum Jülich GmbH, D-52425 Jülich, Germany}

\affiliation{Jülich Centre for Neutron Science JCNS, Forschungszentrum Jülich
GmbH, Outstation at MLZ, Lichtenbergstraße 1, D-85747 Garching, Germany}

\author{W. T. Jin}

\affiliation{Jülich Centre for Neutron Science JCNS and Peter Grünberg Institut
PGI, JARA-FIT, Forschungszentrum Jülich GmbH, D-52425 Jülich, Germany}

\affiliation{Jülich Centre for Neutron Science JCNS, Forschungszentrum Jülich
GmbH, Outstation at MLZ, Lichtenbergstraße 1, D-85747 Garching, Germany}

\author{Y. Xiao}

\affiliation{Jülich Centre for Neutron Science JCNS and Peter Grünberg Institut
PGI, JARA-FIT, Forschungszentrum Jülich GmbH, D-52425 Jülich, Germany}

\author{Y. Su}

\affiliation{Jülich Centre for Neutron Science JCNS, Forschungszentrum Jülich
GmbH, Outstation at MLZ, Lichtenbergstraße 1, D-85747 Garching, Germany}

\author{S. Price}

\affiliation{Jülich Centre for Neutron Science JCNS and Peter Grünberg Institut
PGI, JARA-FIT, Forschungszentrum Jülich GmbH, D-52425 Jülich, Germany}

\author{W. Schmidt}

\affiliation{Jülich Centre for Neutron Science, Forschungszentrum Jülich, Outstation
at Institut Laue-Langevin, BP 156, 38042 Grenoble Cedex 9, France}

\author{K. Schmalzl}

\affiliation{Jülich Centre for Neutron Science, Forschungszentrum Jülich, Outstation
at Institut Laue-Langevin, BP 156, 38042 Grenoble Cedex 9, France}

\author{T. Chatterji}

\affiliation{Institut Laue-Langevin, BP 156, 38042 Grenoble Cedex 9, France}

\author{H. S. Jeevan}

\affiliation{I. Physikalisches Institut, Georg-August-Universität Göttingen, D-37077
Göttingen, Germany}
\affiliation{Department of Physics, PESITM, Sagar Road, 577204 Shimoga, India}

\author{P. Gegenwart}

\affiliation{I. Physikalisches Institut, Georg-August-Universität Göttingen, D-37077
Göttingen, Germany}

\affiliation{Center for Electronic Correlations and Magnetism, Experimental Physics
VI, Universität Augsburg, D-86135 Augsburg, Germany}

\author{Th. Brückel}

\affiliation{Jülich Centre for Neutron Science JCNS and Peter Grünberg Institut
PGI, JARA-FIT, Forschungszentrum Jülich GmbH, D-52425 Jülich, Germany}

\affiliation{Jülich Centre for Neutron Science JCNS, Forschungszentrum Jülich
GmbH, Outstation at MLZ, Lichtenbergstraße 1, D-85747 Garching, Germany}
\begin{abstract}
The magnetic structure of the Eu$^{2+}$ moments in the superconducting
EuFe$_{2}$(As$_{1-x}$P$_{x}$)$_{2}$ sample with \emph{x}\,=\,0.19
has been determined using neutron scattering. We conclude that the
Eu$^{2+}$ moments are aligned along the \textbf{c} direction below
\emph{T}$_{\textup{C}}$ = 19.0(1) K with an ordered moment of 6.6(2)\,$\mu_{\textup{B}}$
in the superconducting state. An impurity phase similar to the underdoped
phase exists within the bulk sample which orders antiferromagnetically
below \emph{T}$_{\textup{N}}$ = 17.0(2) K. We found no indication
of iron magnetic order, nor any incommensurate magnetic order of the
Eu$^{2+}$ moments in the sample.
\end{abstract}

\pacs{74.70.Xa, 75.25.-j, 75.40.Cx}

\maketitle

\section{Introduction}

In the last few years, there have been a flurry of research activity
in the field of unconventional high-\emph{T}$_{\textup{C}}$ superconductivity
\cite{Johnston} due to the discovery of iron-based superconductors
in 2008 \cite{kamihara_08}. Among various classes of Fe-based superconductors
\cite{kamihara_08,Takahashi_08,chen_08,Ren_08,Hsu_08,Guo_10}, the
ternary \textquotedblleft{}122\textquotedblright{} system, \emph{A}Fe$_{2}$As$_{2}$
(\emph{A}\,=\,Ba, Ca, or Sr, etc) with \emph{T}$_{\textup{C}}$
up to 38 K, \cite{rotter_08,Jeevan_08,Sasmal_08}, has been the most
widely studied member of Fe-pnictide superconductors. EuFe$_{2}$As$_{2}$
is an interesting member of the \textquotedblleft{}122\textquotedblright{}
family since the \emph{A} site is occupied by the Eu$^{2+}$, which
is an \emph{S}-state rare-earth ion possessing a 4\emph{f}$^{7}$
electronic configuration with the electron spin \emph{S} = 7/2 \cite{Marchand_1978}.
EuFe$_{2}$As$_{2}$ exhibits a spin density wave (SDW) transition
in the Fe sublattice concomitant with a structural phase transition
at 190 K. In addition, Eu$^{2+}$ moments order in an A-type antiferromagnetic
(AFM) structure at 19 K (ferromagnetic layers ordered antiferromagnetically
along the\emph{ }\textbf{c} direction) \cite{Martin_09,Xiao_09,Xiao_10}.
Superconductivity can be achieved in this system by substituting Eu
with K or Na \cite{Qi_2008, Jeevan_08}, As with P \cite{Ren_09},
and upon application of external pressure \cite{Miclea_09, Terashima_09,Tokiwa_12}.
Doping as well as external pressure lead to a decrease of both the
structural and Fe magnetic phase transition temperatures and eventually
superconductivity appears when both transitions are suppressed enough
\cite{Jeevan_08}. Upon P-doping, the ordering temperature of the
Eu$^{2+}$ moments initially decreases by a few Kelvin until the superconductivity
appears and then increases up to 30 K as the doping is increased further
\cite{Jeevan_11}.

The superconducting dome for the P-doped EuFe$_{2}$As$_{2}$ is quite
narrow compared to the other Fe-based ``122'' pnictides \cite{Nandi_PRL_10}.
It is generally accepted that the Eu$^{2+}$ moments order antiferromagnetically
before the superconducting dome and ferromagnetically after the superconductivity
is suppressed \cite{Jeevan_11}. However, the exact nature of the
Eu$^{2+}$ magnetic order within the superconducting dome remained
controversial \cite{Zapf_13}. Most surprising is the coexistence
of ferromagnetism and superconductivity as recently proposed by many
groups for the P-doped EuFe$_{2}$As$_{2}$ samples \cite{cao_11,Ahmed_10,Nowik,Zapf_11,Zapf_13}.
In particular, Zapf \emph{et al}. \cite{Zapf_11,Zapf_13} concluded
based on macroscopic measurements that the Eu$^{2+}$moments in EuFe$_{2}$(As$_{1-x}$P$_{x}$)$_{2}$
order in a canted A-type antiferromagnetic structure with the spin
component along the \textbf{c} direction being ferromagnetically aligned.
Zapf \emph{et al}. \cite{Zapf_13} also discovered that the A-type
antiferromagnetic order of the Eu$^{2+}$ moments below around 20
K undergoes a spin glass transition at lower temperatures where the
in plane components of the magnetic moments are responsible for the
glassy freezing.

For the superconducting EuFe$_{2}$(As$_{1-x}$P$_{x}$)$_{2}$ samples
with \emph{x}\,=\,0.15, we have recently concluded that the Eu$^{2+}$
moments are primarily aligned ferromagnetically along the \textbf{c}
direction using x-ray resonant magnetic scattering \cite{Nandi_prb_14}.
However, due to the limited sensitivity of the x-ray scattering technique
for the ferromagnetic structures, the moment size could not be determined
from the previous studies. Neutron diffraction studies on a superconducting
Eu(Fe$_{0.82}$Co$_{0.18})$$_{2}$As$_{2}$ single crystal revealed
a long-range ferromagnetic order of the Eu$^{2+}$ moments along the
\textbf{c} direction \cite{Jin_prb_13}. Due to the strong neutron
absorption of the natural Eu (\textasciitilde{} 5800 barns at $\lambda$
= 2.513 Å) together with the small sample mass (\textasciitilde{}\,10
mg) of the P-doped single crystals compared to the Co-doped EuFe$_{2}$As$_{2}$
(\textasciitilde{}100 mg), the magnetic structure determination in
EuFe$_{2}$(As$_{1-x}$P$_{x}$)$_{2}$ via neutron diffraction is
considerably more challenging. The only attempt was made on a powder
sample of the non-superconducting end-member EuFe$_{2}$P$_{2}$ where
it was concluded that the Eu$^{2+}$ moments order ferromagnetically
with a canting angle of 17$^{\circ}$ from the \textbf{c} axis \cite{Ryan_11}.
Based on Mössbauer studies on superconducting polycrystalline samples,
Nowik \emph{et al}. \cite{Nowik} also concluded that the Eu$^{2+}$
moments are aligned ferromagnetically along the \textbf{c} axis with
a possible tilting angle of 20$^{\circ}$ from the \textbf{c} axis
for EuFe$_{2}$(As$_{1-x}$P$_{x}$)$_{2}$ with $x\geq0.2$ . On
the other hand, it was concluded for the Co-doped EuFe$_{2}$As$_{2}$
samples that the canting angle is nearly zero \cite{Jin_prb_13}.
Therefore, it is very important to clarify the moment direction and
the absolute value of the ordered moment using neutron diffraction
for the P-doped EuFe$_{2}$As$_{2}$ single crystals with different
doping levels. Here we report on the neutron scattering studies of
the superconducting EuFe$_{2}$(As$_{1-x}$P$_{x}$)$_{2}$ single
crystal with \emph{x} = 0.19 to explore the details of the magnetic
structure of the Eu$^{2+}$ moments. Our neutron scattering experiments
show that the Eu$^{2+}$ moments order ferromagnetically along the
\textbf{c} direction with an ordered moment of 6.6(2) $\mu_{\textup{B}}$.
No magnetic order associated with Fe as well as no structural phase
transition could be detected.

\section{Experimental Details}

\begin{figure}[t]
\centering
\includegraphics[width=0.5\textwidth]{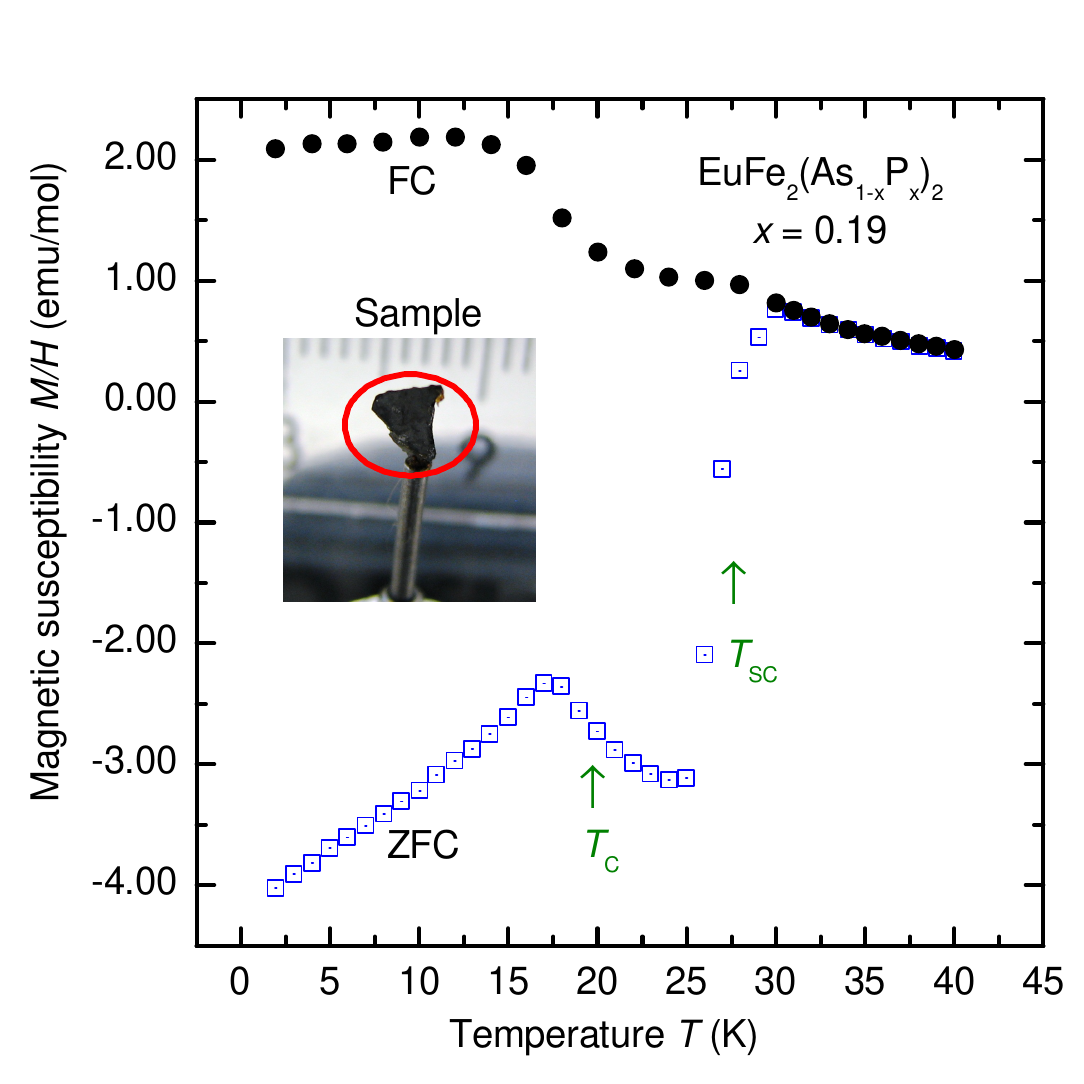}
 \caption{\label{fig1} Temperature dependence of the magnetic susceptibility
measured on heating of the zero-field cooled (ZFC) and field cooled
(FC) sample along the \emph{a-b} plane at an applied magnetic field
of 25 Oe. \emph{T}$_{\textup{\textup{SC}}}$ and \emph{T}$_{\textup{C}}$
denote superconducting and ferromagnetic transition temperatures,
respectively. The transition temperatures were determined from the
minima/maxima of the derivative curve of the ZFC data.}
\end{figure}

Single crystals of EuFe$_{2}$(As$_{1-x}$P$_{x}$)$_{2}$ with $x=0.19$
were grown using FeAs flux \cite{Jeevan_11}. For the scattering measurements,
a 10\,mg as-grown triangular shaped single crystal of approximate
dimensions 2.9\,mm$\times$3.5\,mm$\times3.8$\,mm and thickness
of 0.3 mm was selected (see inset of Fig. \ref{fig1}) and it's phosphorous
content was determined within 1\% accuracy on four different spots
by freshly cleaving the sample using EDX (Energy dispersive x-ray
analysis). High-resolution elastic neutron scattering experiments
were carried out on the cold-neutron triple-axis spectrometer IN12
at the high-flux reactor of the Institut Laue Langevin in Grenoble,
France. Vertically focused Pyrolytic Graphite (PG) (0 0 2) monochromator
and analyzer were used. In addition, monochromator was partially focused
horizontally to optimize flux for the used collimators. To have a
better \emph{q}-space resolution, 60$^{'}$ collimators were used
between the sample and the monochromator as well as between the sample
and the analyzer. A velocity selector in the guide about 39 m upstream
served as a second order filter and for background reduction. The
measurements were carried out with fixed final wave vectors of k$_{f}$
= 2.5 Å$^{-1}$ and 2.85 Å$^{-1}$, which corresponds to neutron wavelengths
of 2.513 Å and 2.205 Å, respectively. Due to the geometrical limitation
of the instrument, the shorter wavelength was employed only for the
measurement of the (2 2 0) reflection. The larger wavelength was used
for the rest of the measurements for improved flux. The single crystal
was mounted on a Vanadium pin using very small amount of GE-Varnish
and mounted inside an ILL orange type cryostat. Furthermore, the sample
stick near the sample as well as the Vanadium pin was covered with
a thick Cd foil for the reduction of background. Slits before and
after the sample were used to further reduce the background. Initially
measurements were performed in the (1 1 0)$_{T}$-(0 0 1)$_{T}$ scattering
plane. Later, the sample was remounted in the (1 0 0)$_{T}$-(0 0
1)$_{T}$ scattering plane for measurements of more magnetic and nuclear
reflections. Measurements at IN12 were performed at temperatures between
2 and 100\,K. For convenience, we will use tetragonal (\emph{T})
notation unless otherwise specified.

\section{Experimental Results}

\subsection{Macroscopic Characterizations}

\begin{figure}[t]
\centering
\includegraphics[width=0.5\textwidth]{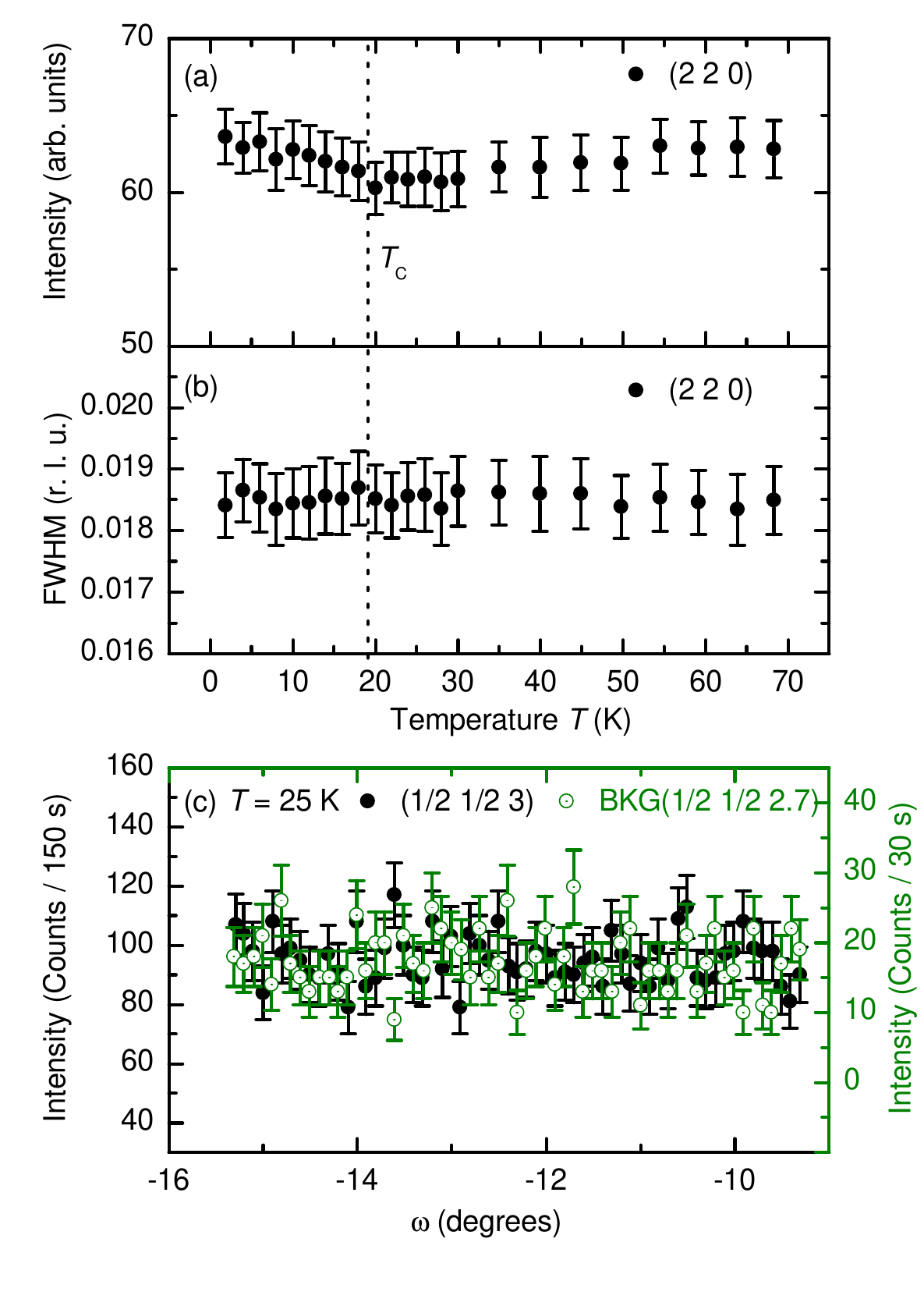}
\caption{\label{fig2}(a) Temperature dependence of the integrated intensity
for the (2 2 0) reflection. (b) Temperature dependence of the Full-Width-Half-Maximum
(FWHM) for the same reflection. (c) Rocking scans at the expected
position of the Fe magnetic order at ($\frac{1}{2}$\,$\frac{1}{2}$\,3)
at 25 K and at the background position of ($\frac{1}{2}$\,$\frac{1}{2}$\,2.7).}
\label{figure2-1-1}
\end{figure}

\begin{figure}[t]
\centering
\includegraphics[width=0.5\textwidth]{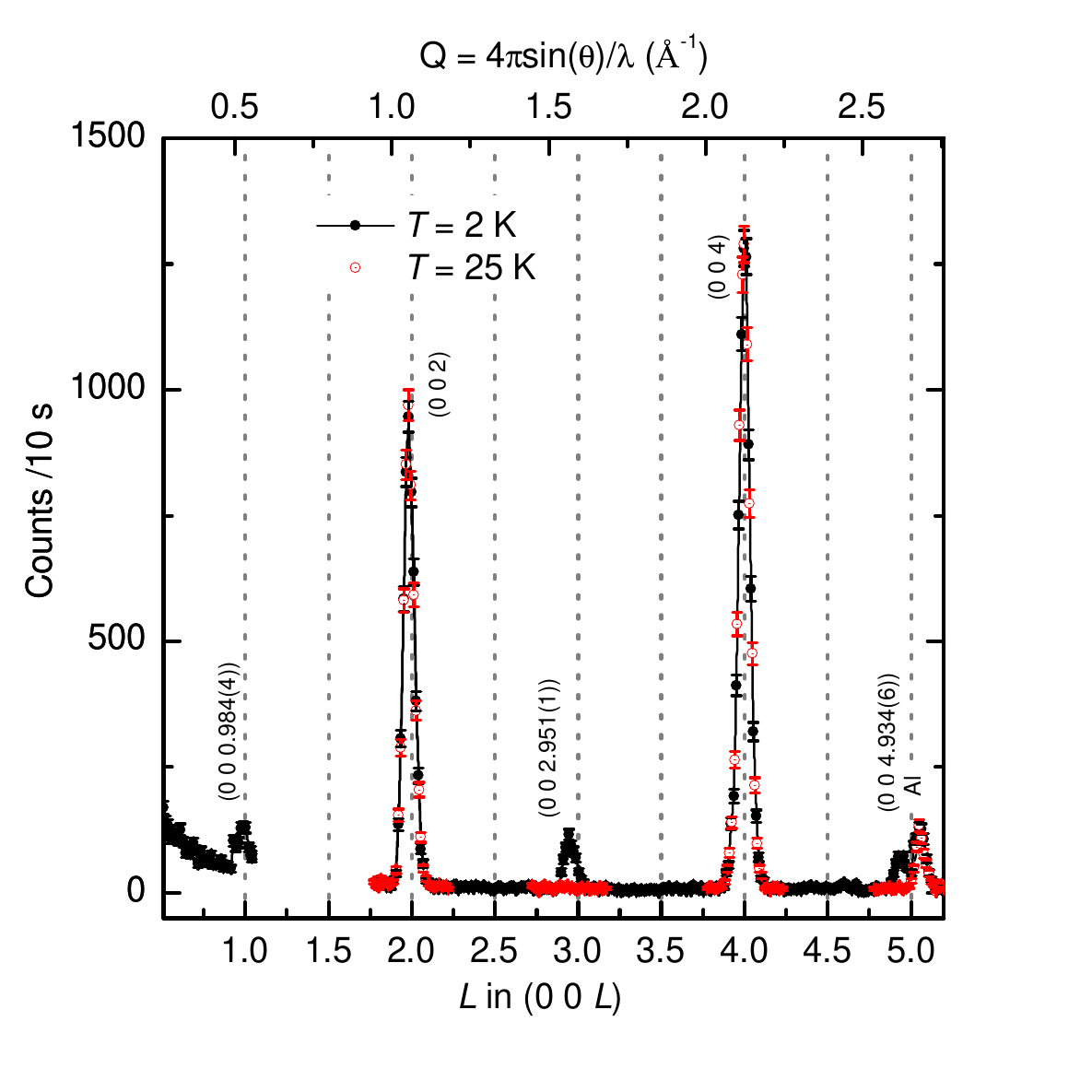}
\caption{\label{fig3} Scans along the {[}0 0\emph{ }1{]} direction at \emph{T}
= 2 and 25 K, respectively. For the (0 0 1) reflection, measurement
has been performed only at 2 K due to the technical difficulties.
Break in measurement data between \emph{L} = 1.1 and 1.75 is due to
the same difficulties. The temperature independent peak near \emph{L}
= 5.04 is due to Aluminum (Al). The peak positions near the (0 0 \emph{L})
with \emph{L} = odd were determined after fitting the peaks using
Gaussian profiles.}
\end{figure}

\begin{figure}[t]
\centering
\includegraphics[width=0.5\textwidth]{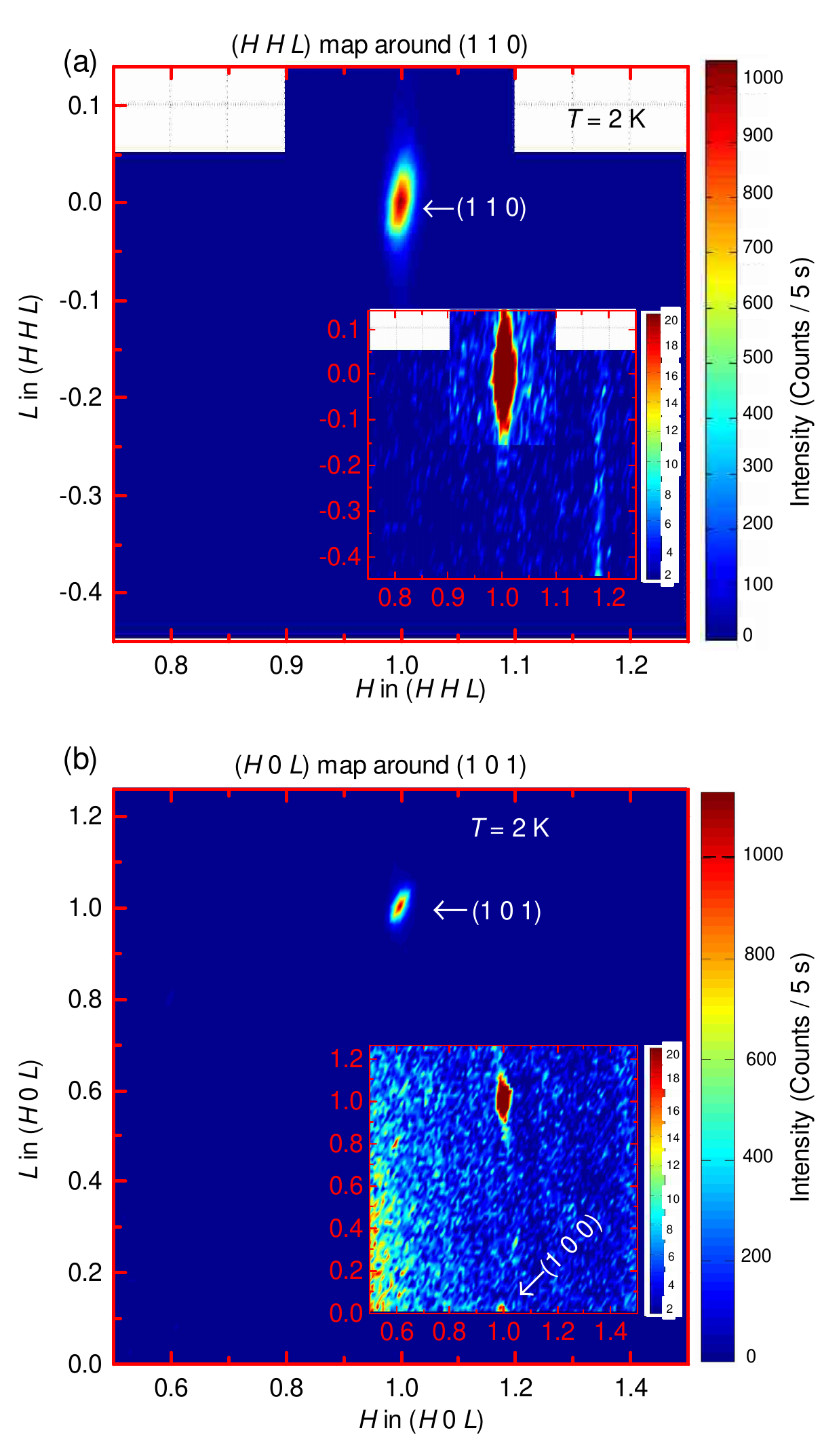}
\caption{\label{fig4}(a) and (b) Two dimensional contour maps in the (\emph{H}
\emph{H} \emph{L}) and in the (\emph{H }0\emph{ L}) planes at \emph{T}
= 2 K. Insets show both figures with different intensity scales to
visualize weakest peaks, if any. No data were collected in the white
rectangular region of Fig. (a). (1 1 0) and (1 0 1) are allowed nuclear
peak positions. The weak peak at (1 0 0) indicates antiferromagnetic
order of the Eu$^{2+}$ moments with propagation vector $\boldsymbol{\tau}$
= (0 0 1) associated with the minor phase.}
\label{figure2}
\end{figure}

Figure\,\ref{fig1} shows temperature dependence of the magnetic
susceptibility \emph{M/H} measured for magnetic field parallel to
the \emph{a-b} plane using a Quantum Design (SQUID) magnetometer.
Zero field cooled magnetization becomes negative at \emph{T}$_{\textup{\textup{SC}}}$
= 27 K, signifying a superconducting transition at this temperature.
At slightly lower temperature and at \emph{T}$_{\textup{C}}$ = 19
K superconducting signal is weakened by the onset of the Eu$^{2+}$
magnetic order and has been observed in heat-capacity measurements
\cite{Jeevan_11}. The transition at \emph{T}$_{\textup{C}}$ is the
ferromagnetic transition of Eu$^{2+}$ moments as supported by the
saturation of magnetization (FC susceptibility) as well as strong
increase of the intensity for the nuclear reflections (see next section)
below this temperature. Furthermore, it was found that the entropy
release associated with this transition is close to the theoretical
value of \emph{R}ln(2\emph{S} + 1) for the Eu$^{2+}$ moments with
spin \emph{S} = 7/2 on similar chemical compositions \cite{Nandi_prb_14,Jeevan_11}.
Superconductivity wins over the Eu$^{2+}$ magnetism if temperature
is lowered further. The results of magnetic susceptibility are consistent
with the published results of Jeevan \emph{et al.} \cite{Jeevan_11}.

\subsection{Neutron diffraction}

\begin{figure}[t]
\centering
\includegraphics[width=0.5\textwidth]{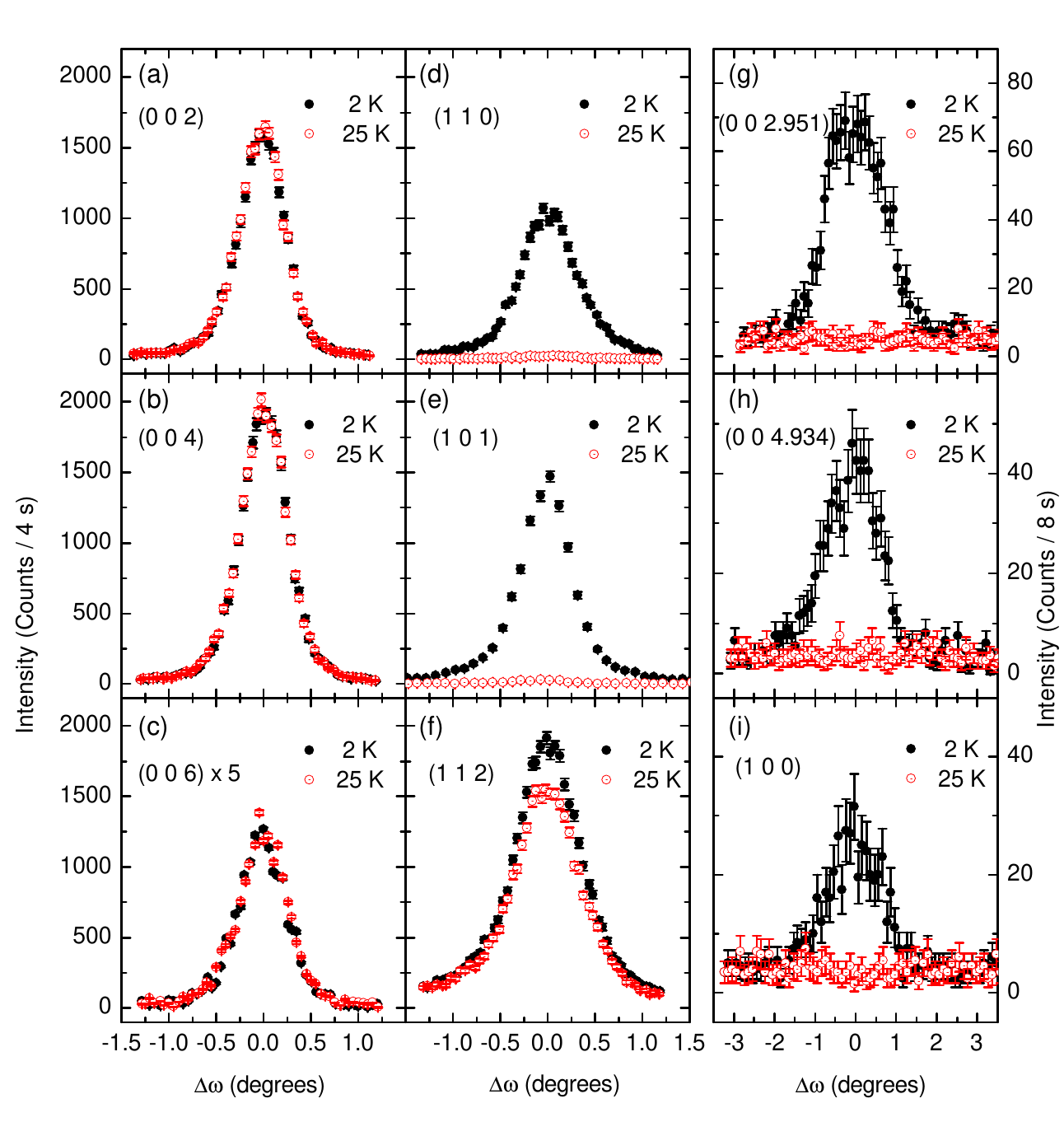}
\caption{\label{fig5}Rocking scans for the (a-c) (0 0 2), (0 0 4) and (0 0
6) reflections, (d-f) (1 1 0), (1 0 1) and (1 1 2) reflections, and
(g-i) (0 0 2.95), (0 0 4.95) and (1 0 0) reflections at 2 and 25 K,
respectively. Note the intensity scales for the (0 0 \emph{L}) reflections
with \emph{L} = odd and \emph{L} = even.}
\end{figure}

It has been established for the parent compound as well as for the
15\% P-doped samples that there is a structural phase transition from
space group $I4/mmm$ to $Fmmm$, with a distortion along the {[}1~1~0{]}
direction at 190 and 49 K, respectively \cite{Xiao_09,Nandi_prb_14}.
To determine whether there is a similar structural phase transition,
($\xi$ $\xi$ 0) scans were performed through the tetragonal (2 2
0) Bragg reflection as a function of temperature. Figure \ref{fig2}(a)
shows temperature dependence of the integrated intensity for the (2
2 0) reflection as the sample was warmed up from 2 to 70 K. It can
be seen that there is an increase of intensity below \emph{T}$_{\textup{C}}$
= 19 K, coincident with the onset of the magnetic order of the Eu$^{2+}$
moments. No change in intensity at higher temperatures associated
with the extinction release \cite{Jin_prb_13} at the structural phase
transition has been observed. Figure \ref{fig2}(b) shows temperature
dependence of the Full-Width-Half-Maximum (FWHM) for the same reflection.
However, the FWHM remains constant within the experimental error in
the investigated temperature range and provides an upper limit of
the structural distortion, $\delta=(a-b)/(a+b)$ $\sim$ 2.5$\times$10$^{-4}$.
For comparison, we note that for the 15\% doped sample a structural
distortion of 5$\times$10$^{-4}$ was observed at \emph{T} = 25 K
\cite{Nandi_prb_14}.

For the undoped parent compound, the Fe moments order magnetically
below 190 K with a propagation vector $\tau=(\frac{1}{2}\,\frac{1}{2}\,1)$
\cite{Xiao_09}. Figure \ref{fig2}(c) shows $\omega$ scans (rocking
curves) through the expected position of the strongest magnetic peak
at $(\frac{1}{2}\,\frac{1}{2}\,3)$ at 25 K and at the background
position of $(\frac{1}{2}\,\frac{1}{2}\,2.7)$. Similar scans were
performed at 2 K and 40 K (not shown). However, we failed to observe
any magnetic peak corresponding to the Fe magnetic order indicating
the absence of the Fe magnetic order within our experimental accuracy.

For the parent compound, Eu$^{2+}$ moments adopt an A-type antiferromagnetic
structure characterized by the propagation vector $\mathbf{\boldsymbol{\tau}}=(0\,0\,1)$.
Figure \ref{fig3} shows Q-scans along the {[}0 0 1{]} direction at
\emph{T} = 2 and 25 K, respectively. Very weak magnetic signals were
observed at \emph{T} = 2 K near (0 0 \emph{L}) with \emph{L} = odd
compared to the allowed nuclear peaks at (0 0 \emph{L}) with \emph{L}
= even. One expects comparable intensities for the (0 0 2) nuclear
peak and the (0 0 3) magnetic peak for the fully ordered moment of
Eu$^{2+}$ as observed for the parent compound \cite{Xiao_09}. The
magnetic peaks are broader in \emph{q}-space compared to the allowed
nuclear peaks and appear at positions incommensurate with the nuclear
peaks. However, this does not ensure that the magnetic structure is
incommensurate with the lattice since the twin satellite peaks are
absent. The observation of magnetic peaks at slightly lower \emph{L}
values (with \emph{L} odd) might be due to the existence of a minor
phase with\textbf{ }\emph{c} lattice parameter larger than the main
phase. The shift of the peak positions of the (0 0 \emph{L}) peaks
with \emph{L} = odd from the nearest integer values increases progressively
with increasing \emph{L} values. This is expected for a minor phase
with a larger \emph{c} lattice parameter. Based on the positions of
the three magnetic peaks a modified \emph{c} lattice parameter, \emph{c}'
= (1.016 $\pm$ 0.003) \emph{c,} can be determined where \emph{c}
corresponds to the lattice parameter of the main phase. It is known
that the decrease in \emph{c} lattice parameter is more pronounced
compared to the \emph{a/b} lattice parameters as a result of chemical
pressure by P-doping \cite{Jeevan_11}. Hence, the peaks near (0 0
\emph{L}) with \emph{L} = odd might be associated with a minor phase
similar to the undoped or underdoped phases. Here we note that we
did not observe (0 0 2-2$\Delta$) and (0 0 4-4$\Delta$) peaks which
is expected for the minor phase ($\Delta$ = 0.016). This might be
due to the weakness of the signal from the minor phase (volume fraction
only 1.5\%, see next section) together with the close proximity of
the lattice parameters with the main phase

Maps in the (\emph{H} \emph{H} \emph{L}) and (\emph{H} 0 \emph{L})
planes were performed at \emph{T} = 2 K to search for additional magnetic
peaks. Figures \ref{fig4}(a) shows a two dimensional contour map
around the very weak nuclear peak (1 1 0). Only the main nuclear peak
was observed, however, with intensity comparable to the strongest
nuclear peaks. Similar contour map around another weak nuclear peak
(1 0 1) in the (\emph{H} 0 \emph{L}) plane was performed and is shown
in Fig. \ref{fig4} (b). In addition to the nuclear peak at (1 0 1),
a very weak magnetic peak was found at (1 0 0) characterized by the
propagation vector $\boldsymbol{\tau}$ = (0 0 1) as observed before.
Very strong intensity at the weakest nuclear peaks and the absence
of incommensurate peaks indicate that the magnetic unit cell might
be the same as the chemical unit cell with magnetic propagation vector
$\boldsymbol{\tau}$ = (0 0 0). Therefore, rocking scans for several
representative nuclear peaks at \emph{T} = 2 and 25 K were performed
to verify if there is any magnetic contribution superimposed on the
nuclear peaks. These scans are shown in Figure \ref{fig5}\,(a-f).
It is clear from the $\omega$ scans of the (0 0 \emph{L}) reflections
with \emph{L} = even in Figs.\,\ref{fig5}\,(a-c) that they have
identical intensities at 2 and 25 K, indicating the absence of magnetic
signal at these positions. However, the weakest nuclear peaks, (1
1 0) and (1 0 1), in Figs. \ref{fig5} (d-e) show very strong magnetic
signals as the intensity of these reflections almost vanishes at 25
K. For the strong nuclear peak (1 1 2), magnetic signal can be observed
on top of the nuclear signal as shown in Figs. \ref{fig5} (f). The
absence of magnetic intensity for the (0 0 \emph{L}) reflections with
\emph{L} even as well as strong intensity at the (1 1 0) position
indicates that the moments are primarily along the \textbf{c} direction.
Figs. \ref{fig5}\,(g-i) show $\omega$ scans of the magnetic peaks
associated with the minor phase. The FWHM of the $\omega$ scans for
the minor phase (\textasciitilde{} 1.5$^{\circ}$) are much larger
than the main phase (\textasciitilde{} 0.5$^{\circ}$) indicating
weak magnetic correlations or incoherently grown minor phase.

\begin{figure}[t]
\centering
\includegraphics[width=0.5\textwidth]{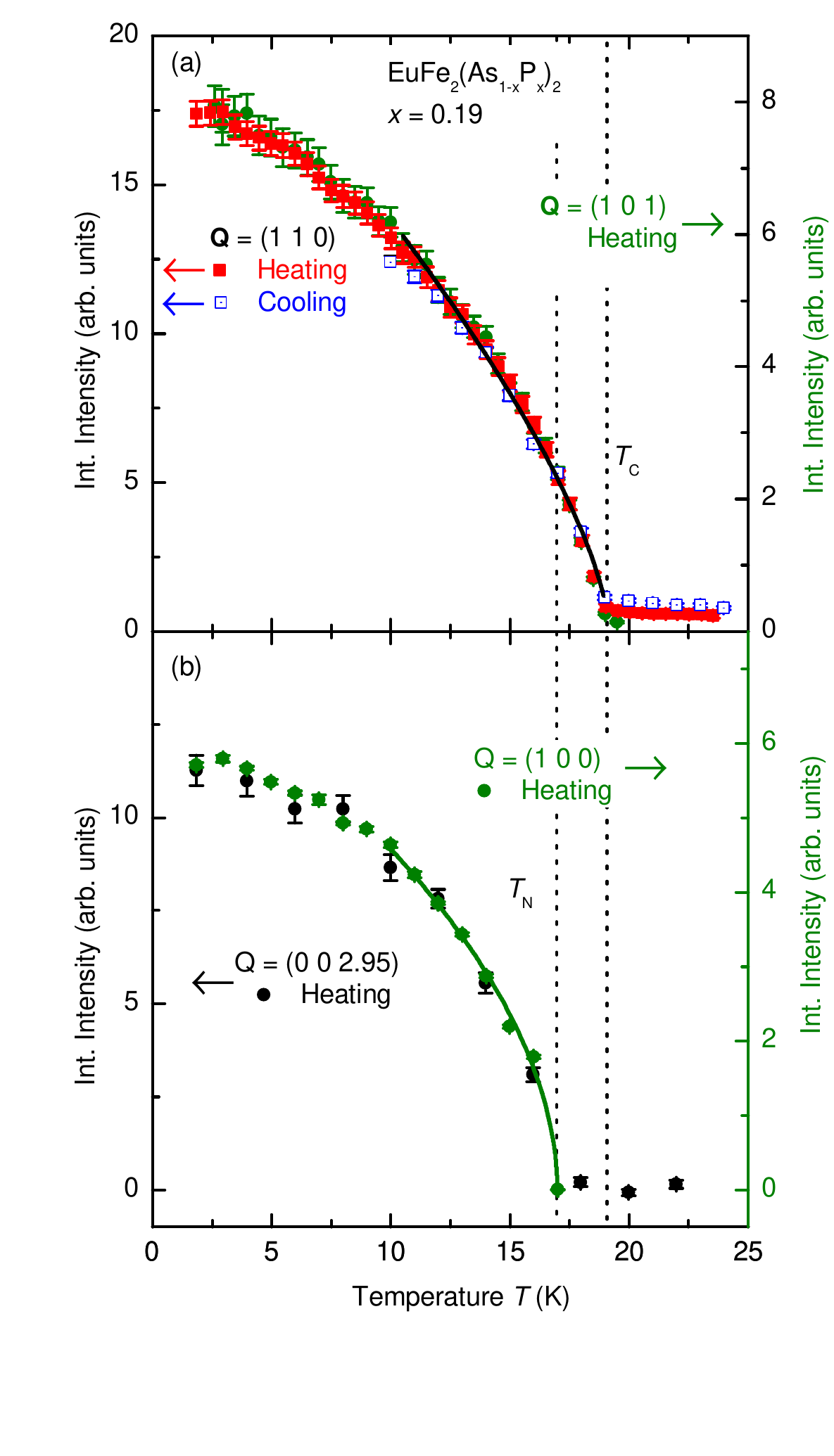}

\caption{\label{fig6}Temperature dependence of the integrated intensity for
the (1 1 0) reflection measured during heating and cooling and for
the (1 0 1) reflection measured during heating. (b) For comparison, we also show the temperature dependencies
of the (0 0 2.95) and (1 0 0) reflections corresponding to the minor
phase. The solid lines are fit to the data as described in the text.}
\end{figure}

Figure \ref{fig6}(a) shows temperature dependencies of the integrated
intensities for the (1 1 0) and (1 0 1) reflections. No hysteresis
was observed between the heating and cooling cycles indicating second
order nature of the phase transition. The integrated intensity ($I\sim m^{2}$,
where \emph{m} is the sublattice magnetization) can be fitted with
a power law of the form $I\thicksim(1-\frac{T}{T_{\textup{C}}})^{2\beta}$
to obtain transition temperature of the Eu$^{2+}$ magnetic order
\emph{T}$_{\textup{C}}$ = 19.0(1)K and an exponent $\beta=0.36(4)$.
Since the (1 0 1) reflection is sensitive to both in and out of plane
magnetic components, the identical nature of the temperature dependencies
for both reflections indicates that the spin canting out of plane,
if any, remains constant over the whole temperature range. Figure \ref{fig6} (b) shows temperature dependencies
of the (0 0 2.95) and (1 0 0) magnetic reflections corresponding to
the antiferromagnetic ordering of the Eu$^{2+}$ moments associated
with the minor phase. Using the same power law fitting to the integrated
intensity, antiferromagnetic ordering temperature, \emph{T}$_{\textup{N}}$
= 17.0(2), and $\beta=0.27(3)$ could be obtained. The exponent for
the main phase is close to that (0.36) (Ref. \onlinecite{kagawa_05})
of the three-dimensional 3D classical Heisenberg model, typical for
rare-earth elements in intermetallic compounds \cite{Brueckel_01,nandi_09}.
Surprisingly, the minor phase order magnetically at a slightly lower
temperature than the major phase, further hinting towards two independent
magnetic phases. A decrease in ordering temperature as has been observed
for the underdoped region of the phase diagram might indicate that
the minor phase is located in the underdoped region in the phase diagram
\cite{Jeevan_11,Zapf_13}. At low temperatures, the intensities of
the reflections from the major and minor phases were fitted using
FULLPROF \cite{Rodrigues} after necessary absorption correction using
DATAP \cite{Coppens}. For the major phase, a ferromagnetic structure
of the Eu$^{2+}$ moments along the \emph{c} direction was assumed
(as concluded in the next section). The intensities of the minor phase
(a total of eight magnetic reflections) were fitted with an A-type
antiferromagnetic structure of the Eu$^{2+}$ moments \cite{Xiao_09}
with the moment size same as that of the major phase. Comparing the
scale factors for the major phase and minor phase, we estimate the
volume fraction of the minor phase to be (1.5 $\pm$ 0.4)\,\% .

\subsection{Magnetic structure of the Eu$^{2+}$moments}

\begin{figure}[t]
\centering
\includegraphics[width=0.5\textwidth]{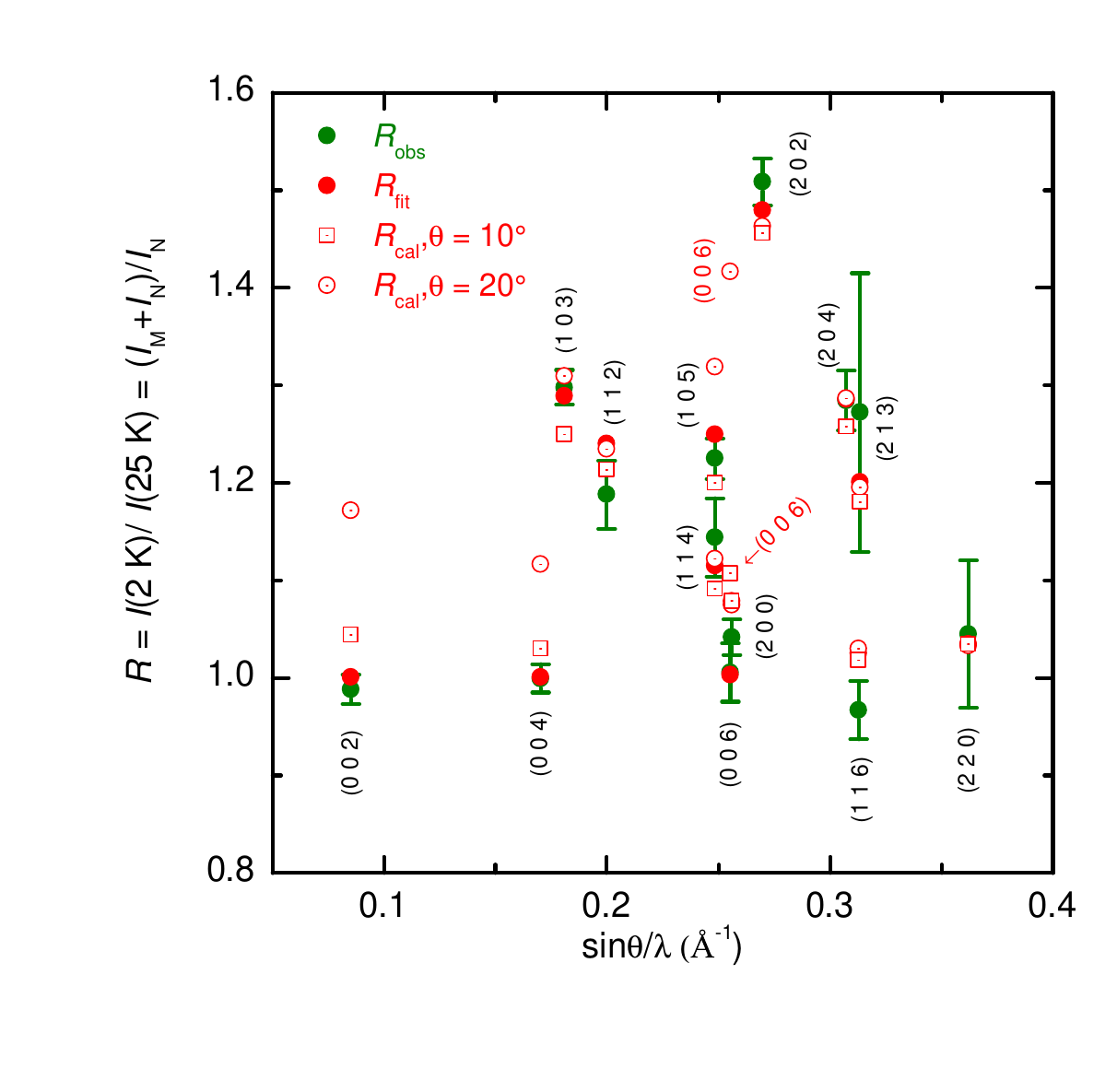}

\caption{\label{fig7}Comparison between the observed ratio (\emph{R}$_{obs}$),
fitted ratio (\emph{R}$_{fit}$), and calculated ratio (\emph{R}$_{cal}$)
for the measured reflections. Details have been outlined in the text. }
\end{figure}

We now turn to the determination of the magnetic moment configuration
for the Eu$^{2+}$ moments in the major phase. Here we note that only
ferromagnetic structures with magnetic moments along the three crystallographic
directions \textbf{a}\emph{, }\textbf{b}\emph{, }\textbf{c} are allowed
by symmetry. No antiferromagnetic structure with $\mathbf{\boldsymbol{\tau}}$\,=\,(0\,0\,0)
is possible in this case for symmetry reasons. Since for magnetic
neutron scattering, the scattered intensity is sensitive to the component
of the magnetic moment perpendicular to \textbf{Q}, absence of magnetic
intensity for the nuclear (0 0 \emph{L}) reflections (see Figs. \ref{fig5}
(a-c)) indicates that the moments are primarily along the \textbf{c}
direction. However, to put an upper limit on the canting angle as
well as to determine the moment size, a magnetic structure refinement
is needed. Conventional magnetic structure refinement using integrated
intensity is difficult in this case because, (a) the number of measured
reflections is limited due to the use of thermal neutrons in a triple
axis spectrometer and, (b) the errors associated with the resolution
effects and absorption corrections in determining absolute integrated
intensities. In particular, absorption correction is difficult with
an irregular shaped sample as shown in inset to the Fig.\,\ref{fig1}
.To circumvent these problem, we have refined the magnetic structure
using the ratios of intensities between 2 and 25 K, \emph{i.e} using\emph{
R} = \emph{I}(2 K)/\emph{I}(25 K) = 1 +\emph{ I}$_{\textup{M}}$/\emph{I}$_{\textup{N}}$,
assuming the nuclear intensity, \emph{I}$_{\textup{N}}$, does not
vary between 2 and 25 K. By using the ratios of intensities, one can
exclude the artifacts due to the resolution effects in a triple axis
spectrometers as well as absorption corrections. The intensity ratios
measured for equivalent reflections and for repeated measurements
of the same reflection were averaged together to give a mean value
of the ratio and is shown in Fig. \ref{fig7} together with the calculated
ratios. The procedure for calculating intensity ratios is outlined
below and has been implemented using MATHEMATICA \cite{Mathematica}.

\begin{table}
\caption{Parameters used for least square refinements of the ratios \cite{Xiao_09,Ryan_11}.
Occupancies corresponding to the As and P sites were fixed according
to the chemical composition.}

\begin{ruledtabular} %
\begin{tabular}{ccccccc}
Atom  & \multicolumn{4}{c}{Position in \emph{I}4/\emph{mmm}} & B(Å\textbf{$^{2}$)} & \emph{n}\tabularnewline[\doublerulesep]
\noalign{\vskip\doublerulesep}
 & site  & \emph{x}  & \emph{y} & \emph{z} & \multicolumn{2}{c}{}\tabularnewline[\doublerulesep]
\hline
\noalign{\vskip\doublerulesep}
Eu  & 2\emph{a}  & 0  & 0  & 0 & 0.8 & 1.00\tabularnewline
Fe & 4\emph{d}  & $\frac{1}{2}$ & 0  & $\frac{1}{4}$ & 0.3 & 1.00\tabularnewline
As & 4\emph{e}  & 0  & 0  & 0.3615 & 0.3 & 0.81 \tabularnewline[\doublerulesep]
\multirow{1}{*}{P} & 4\emph{e} & 0 & 0 & 0.3615 & 0.3 & 0.19\tabularnewline
\multicolumn{7}{c}{\emph{a} = 3.91 (2) Å, \emph{c} = 11.77(4) Å}\tabularnewline
\end{tabular}\end{ruledtabular} \label{structural_parameters}
\end{table}

\begin{table*}
\caption{The observed, fitted and calculated ratios for the reflections with
magnetic contribution of EuFe$_{2}$(As$_{0.81}$P$_{0.19}$)$_{2}$.
Various cases have been considered as described in the footnote and
in the main text.}

\begin{ruledtabular} %
\begin{tabular}{ccrcccccccc}
\multicolumn{1}{c}{} &  &  & $\sin$$\theta$/$\lambda$ & \emph{R$_{obs}$} & \emph{R$_{fit}$$^{a}$} & \emph{R$_{fit}$$^{b}$} & \emph{R$_{fit}$$^{c}$} & \emph{R$_{fit}$$^{d}$} & \emph{R}$_{cal}$\emph{$^{e}$} & \emph{R$_{cal}{}^{f}$}\tabularnewline
\multicolumn{1}{c}{\emph{h}} & \emph{k}  & \emph{l} & Å$^{-1}$ & $\times10^{2}$ & $\times10^{2}$ & $\times10^{2}$ & $\times10^{2}$ & $\times10^{2}$ & $\times10^{2}$ & $\times10^{2}$\tabularnewline
\hline
 0 & 0  & 2  & 0.0851 & 98.8$\pm$ 1.4  & 100.1 & 100.0 & 104.2 & 111.7 & 104.4 & 117.1\tabularnewline
0 & 0  & 4  & 0.1702 & 99.9$\pm$1.4 & 100.1 & 100.0 & 102.9 & 107.9 & 103.0 & 111.7\tabularnewline
0 & 0  & 6  & 0.2553 & 100.6$\pm$ 2.4 & 100.3 & 100.0 & 110.2 & 128.5 & 110.7 & 141.7\tabularnewline
1 & 1  & 2  & 0.2000 & 118.8$\pm$ 3.5 & 124.0 & 124.0 & 122.9 & 116.0 & 124.1 & 123.5\tabularnewline
1 & 1  & 4  & 0.2485 & 114.4$\pm$ 4.4 & 111.5 & 111.5 & 111.2 & 108.3 & 111.7 & 112.2\tabularnewline
1 & 1  & 6  & 0.3129 & 96.7$\pm$ 3.0 & 102.5 & 102.5 & 102.5 & 102.0 & 102.6 & 102.9\tabularnewline
2 & 1 & 3  & 0.3134 & 127.2$\pm$ 14.0 & 120.1 & 120.0 & 119.1 & 113.3 & 120.1 & 119.6\tabularnewline
2 & 2 & 0  & 0.3620 & 104.5$\pm$ 7.6  & 103.4 & 103.4 & 103.2 & 102.2 & 103.4 & 103.3\tabularnewline
1 & 0  & 3  & 0.1808 & 129.8$\pm$ 1.8 & 128.9 & 128.9 & 128.3 & 121.1 & 129.7 & 130.9\tabularnewline
1 & 0  & 5  & 0.2483 & 122.5$\pm$ 2.0 & 124.9 & 124.9 & 125.7 & 121.8 & 126.9 & 131.9\tabularnewline
2 & 0  & 0  & 0.2559 & 104.2$\pm$ 1.8 & 107.8 & 107.8 & 107.4 & 105.1 & 107.8 & 107.5\tabularnewline
2 & 0  & 2  & 0.2698 & 150.9$\pm$ 2.4 & 147.9 & 148.0 & 145.6 & 131.6 & 147.9 & 146.3\tabularnewline
2 & 0 & 4 & 0.3074 & 128.4$\pm$ 3.0 & 128.5 & 128.5 & 127.4 & 119.6 & 128.8 & 128.7\tabularnewline
\hline
\multicolumn{5}{c}{Parameters:} & \emph{m}$_{\textup{Eu}}$ = 6.6(2) $\mu_{\textup{B}}$ & \emph{m}$_{\textup{Eu}}$ = 6.6(2) $\mu_{\textup{B}}$ & \emph{m}$_{\textup{Eu}}$ = 6.5(2) $\mu_{\textup{B}}$ & \emph{m}$_{\textup{Eu}}$ = 5.5(5) $\mu_{\textup{B}}$ & \emph{m}$_{\textup{Eu}}$ = 6.6 $\mu_{\textup{B}}$ & \emph{m}$_{\textup{Eu}}$ = 6.6 $\mu_{\textup{B}}$\tabularnewline
\multicolumn{5}{c}{} & $\theta=$ 2(6)$^{\circ}$ & $\theta=$ 0$^{\circ}$ & $\theta=$ 10$^{\circ}$ & $\theta=$ 20$^{\circ}$ & $\theta=$ 10$^{\circ}$ & $\theta=$ 20$^{\circ}$\tabularnewline
\multicolumn{5}{c}{Quality factors:} & \emph{R}$_{wp}$ = 3.3\% & \emph{R}$_{wp}$ = 3.2\% & \emph{R}$_{wp}$= 6.7\% & \emph{R}$_{wp}$ = 15.8\% &  & \tabularnewline
 &  &  &  &  & $\chi_{R}^{2}$ = 1.34 & $\chi_{R}^{2}$  = 1.22 & $\chi_{R}^{2}$ = 3.8 & $\chi_{R}^{2}$ = 25.0 &  & \tabularnewline
\hline
\multicolumn{11}{c}{$^{a}$Fitted with \emph{m}$_{\textup{Eu}}$ and canting angle $\theta$,
$^{b,\, c,\, d}$ Fitted with \emph{m}$_{\textup{Eu}}$ and canting
angle was fixed to 0$^{\circ}$, 10$^{\circ}$, 20$^{\circ}$, respectively. }\tabularnewline
\multicolumn{11}{c}{$^{e}$Calculated with \emph{m}$_{\textup{Eu}}$ = 6.6 $\mu_{\textup{B}}$
and $\theta$ = 10$^{\circ}$, $^{f}$Calculated with \emph{m}$_{\textup{Eu}}$
= 6.6 $\mu_{\textup{B}}$ and $\theta$ = 20$^{\circ}$.}\tabularnewline
\multicolumn{11}{c}{}\tabularnewline
\hline
 & \multicolumn{10}{c}{$R_{wp}=100\frac{\sum_{n}w_{n}\left|I_{obs,n}^{2}-I_{calc,n}^{2}\right|}{\sum_{n}w_{n}I_{obs,n}^{2}}$.\emph{
}$w_{n}=1/\sigma_{n}^{2}$, is the weight where $\sigma_{n}^{2}$
is the variance of $I_{obs,n}$.}\tabularnewline
\multicolumn{11}{c}{$\chi_{R}^{2}=\frac{1}{N-p}\sum_{n}w_{n}(I_{obs,n}-I_{calc,n})^{2}$.
\emph{p }is the number of parameters, \emph{N} is the total number
of observations.}\tabularnewline
\end{tabular}\end{ruledtabular}

\label{calcualted ratios}
\end{table*}
The cross-section for the magnetic neutron scattering can be written
as \cite{Squires_book}:
\begin{eqnarray}
(\frac{d\sigma}{d\Omega})_{el}^{M} & = & (\gamma r_{0})^{2}N_{M}\frac{(2\pi)^{3}}{v_{0m}}\sum_{\vec{\tau}_{m}}\left|F_{m}(\vec{\tau}_{m})\right|^{2}\nonumber \\
 &  & \times\{1-(\vec{\tau}_{m}\cdot\hat{\eta})_{av}^{2}\}\delta(\vec{k}-\vec{\tau}_{m})\label{form factor}
\end{eqnarray}
where
\[
F_{m}(\vec{\tau}_{m})=\frac{1}{2}g\left\langle \vec{S}^{\eta}\right\rangle f(\vec{\tau}_{m})\sum_{d}\sigma_{d}\exp(i\vec{\tau}_{m}\cdot\vec{d})\exp(-w_{d})
\]

Here, $\gamma r_{0}=5.36\times10^{-15}$\,m, N$_{M}$ = Number of
magnetic unit cell in the crystal, $v_{0m}$ is the volume of the
magnetic unit cell, $\vec{\tau}_{m}$ is the vector in magnetic reciprocal
lattice, $w_{d}$ is the Debye-Waller Factor (DWF). $\hat{\eta}$
is the direction of spin $\vec{S}$ where \emph{``av''} means average
over domains %
\footnote{In the calculation four possible orientational domains, (u v w), (-u
v w), (u -v w) and (-u -v w),  were considered, %
}. \emph{g} is the Landé g-factor and $f(\vec{\tau}_{m})$ is the magnetic
form factor at the scattering vector $\vec{\tau}_{m}$. $\sigma_{d}$
= 1 in this case and \emph{d} denotes magnetic atoms in the magnetic
unit cell.

Similarly, the coherent elastic nuclear scattering cross-section can
be written as:
\begin{equation}
(\frac{d\sigma}{d\Omega})_{elastic}^{N}=N_{N}\frac{(2\pi)^{3}}{v_{0}}\sum_{\vec{\tau}_{m}}\left|F_{N}(\vec{k})\right|^{2}\delta(\vec{k}-\vec{\tau})\label{eq:2, nuclear}
\end{equation}

\[
F_{N}(\vec{k})=\sum_{d}\bar{b}_{d}\exp(i\vec{k}\cdot\vec{d})\exp(-w_{d})
\]

where symbols have usual meaning similar to magnetic scattering and
$\bar{b}_{d}$ is the nuclear scattering length for the \emph{d}$^{th}$
ion. The integrated intensity of a reflection at a scattering angle
$2\theta$ can be written as:

\begin{equation}
I=\frac{V}{v_{0}^{2}}\phi\lambda^{3}\times A(\theta)\times B(\mu)\times y_{s}\times\frac{d\sigma}{d\Omega}\label{eq:intensity}
\end{equation}

where \emph{V} is the volume of the crystal, $\phi$ is the incident
neutron flux, \emph{A}($\theta$) contains angular dependent factors
such as Lorentz factor, resolution factor and \emph{B}($\mu$) is
the absorption correction factor. The secondary extinction correction
factor, \emph{y}$_{s}$, depends on the mosaic width, scattering angle
and average scattering cross-section for a particular reflection.
It can be seen from Eqns. (1-3) that in the calculated ratio all factors
except, (i) absolute value of the moment, (ii) moment direction, (iii)
DWF and, (iv) extinction correction cancels out. However, in conventional
single crystal neutron diffraction experiment on similar compounds
it was found that the extinction corrections are small and we will
neglect them in the subsequent analysis. Fitting the nuclear intensities
at \emph{T} = 25 K with and without an extinction parameter (not shown)
clearly demonstrates that the extinction can be neglected. In fact,
adding an extinction parameter worsen the quality of fit. For the
accurate determination of the DWF, a conventional neutron diffraction
experiment up to high \textbf{Q} is needed. We have used the DWF's
for the parent compound \cite{Xiao_09} and used same DWF for P as
for As. Furthermore, we have compared our results without applying
any DWF and the results are the same within errors. For the estimation
of the nuclear intensity, the \emph{z} position of As, \emph{z}(As),
is required. Since the \emph{z}(As) varies from 0.363 for the parent
compound to 0.360 for the end-member, we have used 0.3615 for the
calculation of the nuclear intensity. All the structural parameters
used in the refinement are listed in Table \ref{structural_parameters}.
We have excluded in the refinement the very weak nuclear reflections
with very strong magnetic contributions such as (1 1 0), (1 0 1) and
(2 1 1) since the intensity of these reflections slowly decreases above \emph{T}$_{\textup{C}}$. The
resulting fitted ratios are shown along with observed ratios in Fig.\,\ref{fig7}.
We obtained magnetic moment size of Eu$^{2+}$, \emph{m}$_{\textup{Eu}}$
= 6.6(2) $\mu_{\textup{B}}$ with a possible canting angle $\theta$
= 2(6)$^{\circ}$ with the aid of MATHEMATICA \cite{Mathematica}.
The magnetic moment size is close to the theoretically predicted value
of 7 $\mu_{\textup{B}}$ for a $J=\frac{7}{2}$ Eu$^{2+}$ ion.

Here we note that the canting angle of 2$^{\circ}$ from the fitting
of the ratios is much smaller than the reported values of Ryan \emph{et
al}. \cite{Ryan_11} as well as Nowik \emph{et al}. \cite{Nowik}
of approximately 20$^{\circ}$. Therefore, we have considered five
different scenarios, namely, (b-d) fitted the ratios with canting
angle fixed to 0, 10 and 20$^{\circ}$, respectively and (e-f) simulated
the ratios with moment size fixed to 6.6\,$\mu_{\textup{B}}$ and
two different canting angles, 10 and 20$^{\circ}$, respectively.
The simulated results are shown in Fig. \ref{fig7} and both the fitted
and simulated results are summarized in Table \ref{calcualted ratios}.
It can be clearly seen that the (0 0 \emph{L}) reflections deviate
strongly from the observed ratio if any canting angle is added. The
deviation is strongest for the (0 0 6) reflection. Therefore, measurement
of accurate integrated intensities for weak reflections such as (0
0 6) using a triple axis spectrometer provide strongest constraint
on the canting angle. Fitting with any canting angle is poor as can
be seen from the obtained ratios as well as agreement factors in Table
\ref{calcualted ratios}. Fitting without any canting angle ($\theta=0^{\circ}$)
produces the best $\chi_{R}^{2}$ for the calculated ratios. Therefore,
we conclude that the canting angle is zero in the present case, \emph{i.e.}
moments are aligned along the \textbf{c }direction.

\section{Conclusion}

In conclusion, the magnetic structure of the Eu$^{2+}$ moments in
superconducting EuFe$_{2}$(As$_{1-x}$P$_{x}$)$_{2}$ sample with
$x=0.19$ has been determined using neutron scattering. We conclude
that the Eu$^{2+}$ moments order along the \textbf{c} direction below
\emph{T}$_{\textup{C}}$ = 19.0(1) K with an ordered magnetic moment
of 6.6(2) $\mu_{\textup{B}}$ in the superconducting state. An impurity
phase similar to the underdoped phase exists within the bulk sample
which orders antiferromagnetically below \emph{T}$_{\textup{N}}$
= 17.0(2) K. Further measurements are necessary to elucidate the exact
nature of the minor phase. We found no indication of iron magnetic
order, nor an incommensurate magnetic order of the Eu moments associated
with the major phase. The proposed canted antiferromagnetic order
could not be detected in the superconducting sample. It will be interesting
to investigate Eu magnetic order for higher doping level in EuFe$_{2}$(As$_{1-x}$P$_{x}$)$_{2}$
system using single crystal neutron diffraction.

\bibliographystyle{apsrev} \bibliographystyle{apsrev}
\begin{acknowledgments}
S. N. likes to acknowledge S. Zapf and M. Dressel for fruitful discussion.
Work at Göttingen was supported by the German Science Foundation through
SPP 1458.\bibliographystyle{EuFeAs}
\bibliography{EuFeAs}
\end{acknowledgments}

\end{document}